\documentstyle[multicol,prl,aps,epsf,psfig,epsfig]{revtex}

\begin{document}
\sloppy

\title{Free-energy Barrier to Melting of Single-chain Polymer Crystallite}
  \author{Wenbing Hu$^1$, Daan Frenkel$^1$, Vincent B. F. Mathot$^2$\\$^1$FOM
Institute for Atomic and Molecular Physics,\\ Kruislaan 407, 1098 SJ Amsterdam,
The Netherlands \\$^2$DSM Research, P. O. Box 18 Geleen, The Netherlands}

\maketitle

\begin{abstract}
We report Monte Carlo simulations of the melting of a
single-polymer crystallite.  We find that, unlike most atomic and
molecular crystals, such crystallites can be heated appreciably
above their melting temperature before they transform to the
disordered "coil" state.  The surface of the superheated
crystallite is found to be disordered.  The thickness of the
disordered layer increases with superheating.  However, the
order-disorder transition is not gradual but sudden. Free-energy
calculations reveal the presence of a large free-energy barrier to
melting.
\end{abstract}

\begin{multicols}{2}

\begin{figure}[h]
\centering\epsfig{file=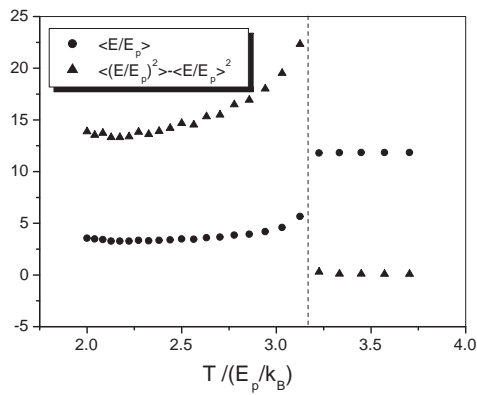,
         height=6 cm}
\caption{ Stepwise heating curve of the potential energy $<E/E_p>$ (spheres) and
its fluctuations (triangles) for a single-chain crystallite of a lattice polymer
with $1024$ units.  The system is allowed to relax during $10^6$ MC cycles at
each temperature.  The temperature was increased by decreasing $E_p/(k_BT)$ with
a fixed step size of $0.01$.  Another $10^6$ MC cycles were used to compute
average properties at each temperature. The dashed line indicates the
approximate temperature where irreversible melting took place in the
simulations.
}
\label{fig1}
\end{figure}

It is easy to supercool liquids but difficult to superheat solids.
The reason is that the surface of a crystal can melt at a
temperature that is well below the bulk melting temperature.
\cite{1}  As a consequence, solids usually melt from the surface
inward without significant superheating.  There exist experimental
studies that show superheating of solids, but in these experiments
the crystals are either confined in a  non-melting matrix \cite{2}
or the experiments reveal superheating of one particular crystal
surface only. \cite{3}
\\  In the present paper, we consider the melting of a macromolecular crystal.
The freezing of polymer crystals has been the subject of much
experimental and theoretical work, but much less attention has
been paid to the melting of polymer crystallites. \cite{4,5}
 This is also true for the case of a single-chain polymer
crystallite. In fact, an extensive numerical study of the freezing
of such a crystal has been reported by Muthukumar and
coworkers~\cite{WM,LM}and by Fukui et al.~\cite{F}. However, there
is no corresponding analysis of the melting of single-chain
crystals, in spite of the fact that this is the simplest example
of polymer-crystal melting. The freezing and melting of
single-polymer crystallites is expected to differ from the
corresponding phenomenon in systems of small molecules. The reason
is that the size of a single-chain crystallite is limited by the
length of the polymer chain.  This implies that, even if such a
crystal would have its equilibrium morphology, it would melt well
below the bulk melting temperature. Moreover, when a single
polymer partially melts (or dissolves), then the molten monomers
cannot escape from the surroundings of the crystallite. Rather
they stay around as a "corona" and can, in this way, affect the
remainder of the melting process. The simulations presented below
show that these features make the melting of polymer crystallites
qualitatively different from that of atomic or molecular crystals.
 \\ To study the melting of a single-chain crystallite, we used a lattice
model. Lattice models provide a highly simplified picture of
freezing and melting. Nevertheless, it has been shown~\cite{TT}
that such models are sufficiently flexible to account for the
phenomenon of surface melting in simple "atomic" systems. The
polymer lattice model that we used in the present study is
described in Ref. \cite{6}. In this model, polymers live on a
simple cubic lattice, but the polymer bonds can be directed both
along main axes of the lattice and along the face and body
diagonals: 26 directions in all.  The polymers can be
semi-flexible and have an attractive nearest-neighbor interaction.
This interaction is anisotropic: parallel polymer bonds attract
more strongly than non-parallel bonds.  Increasing the anisotropy
of the polymer-polymer interaction stabilizes the crystalline
state with respect to both the molten-globule ("liquid") state and
the coil ("vapor") state. \cite{6} In its most general
formulation, this  model is characterized by three energy
parameters. $E_c$ is the energy required to make a kink in the
polymer.  The larger $E_c$, the stronger the tendency of the
polymers to be linear.  All non- collinear bonds are assumed to
have the same energetic cost.  The interaction between two
non-bonded adjacent bonds is controlled by two parameters: the
isotropic interaction energy $B$, which measures the energy change
when forming one polymer-solvent contact from polymer-polymer and
solvent-solvent contacts,and the closely related anisotropic
interaction energy $E_p$, which only acts between parallel polymer
bonds. By varying $E_c$, $B$ and $E_p$ we can change the
"phase-diagram" of the polymer. A flexible ($E_c = 0$) polymer
with a large $B$ undergoes a coil-globule transition. 
In contrast, a polymer with a large value of $E_p$ but
small $B$ will go directly from the coil state to the crystalline
state - even when $E_c=0$.  In what follows, we consider the
simplest model of a crystallizing polymer, namely one where $E_c$
and $B$ vanish, but $E_p$ does not.  At first sight, it would seem
that such a fully flexible polymer cannot form an ordered crystal.
However, the interaction energy $E_p$ has the effect to align
polymers in the crystalline state.  We performed simulations of a
single-chain crystallite consisting of $1024$ monomers.  To
minimize finite-size effects, we used periodic boundary conditions
and a simulation box containing $256^3$ lattice points.  This was
large enough to avoid interactions of the polymer with its
periodic image, even in the coil state.
\\  Initially, we prepared the polymer chain in a rectangular crystal with a
folding length of $4$ monomers and lateral dimensions of $16$ by $16$ lattice
units at a temperature $T=2.0E_p/k_B$.  Subsequently, this crystallite was
annealed for $10^6$ MC cycles, where one cycle corresponds to one trial move per
monomer.  This annealing results initially in a thickening of the crystallite.
Longer annealing does not result in further changes of the crystallite shape.
This suggests that, at the end of the annealing run, the crystallite has reached
its equilibrium morphology.  Having thus prepared an equilibrated single-chain
crystal, we slowly heat it.  To this end, we decrease the ratio $E_p/(k_BT)$ by
$0.01$ every $10^6$ MC cycles.  Figure \ref{fig1} shows the average potential
energy $<E/E_p>$ of the single chain as a function of temperature.  We choose
the potential energy of a fully ordered bulk crystal as our zero of energy.  As
can be seen in figure \ref{fig1}, there is a sudden change in $<E/E_p>$ around
$3.125 E_p/k_B$.  This change is due to the sudden and irreversible melting of
the compact crystallite to an expanded coil state.  This transition to the coil
state is preceded by an appreciable pre-transitional increase of the energy.
Figure \ref{fig1} also shows the temperature dependence of the fluctuations in
the internal energy.  As is expected, the pretransitional rise of the internal
energy is accompanied by an increase in the energy fluctuations.

\begin{figure}[h]
\centering\epsfig{file=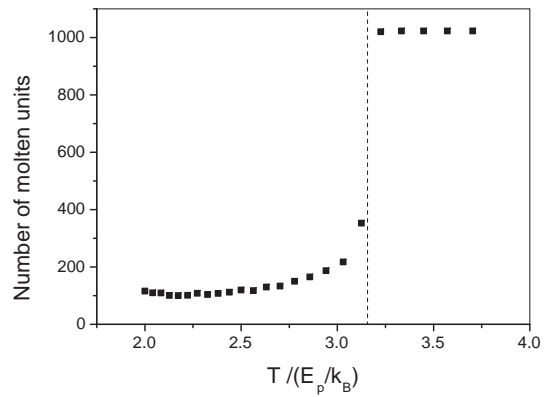,
         height=6 cm}
\caption{Stepwise heating curve of the number of molten units (see text) under
the same conditions as shown in figure \ref{fig1}.
}
\label{fig2}
\end{figure}

  In order to distinguish between bonds that belong to the crystalline core
and those that belong to the corona, we need to have an order parameter that is
sensitive to the environment of a bond.  We found that bonds inside a
crystallite, always have more than $5$ parallel neighbors, while the
overwhelming majority of bonds in the coil state have less.  We therefore refer
to bonds with fewer than $5$ parallel neighbors as "molten units".  In figure
\ref{fig2}, we show the dependence of the number of molten units on temperature.
At low temperatures, about $100$ of the units that are identified as "molten",
can be attributed to the bonds on the hairpin folds at the crystallite surface.
When the crystallite is heated close to the onset of abrupt melting, the number
of molten units increases significantly.

\begin{figure}[h]
\centering{\textbf{A.}\epsfig{file=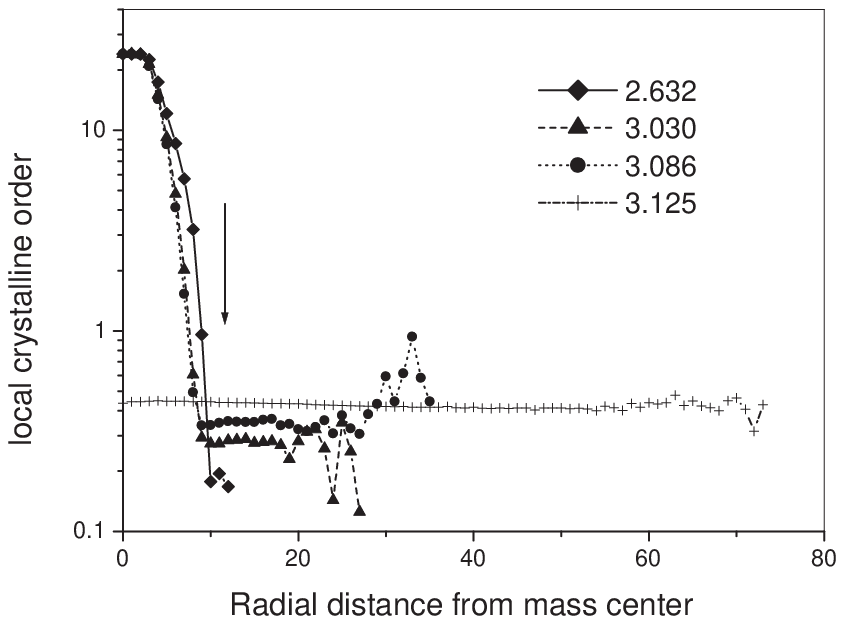,height=6 cm}
           \textbf{B.}\psfig{file=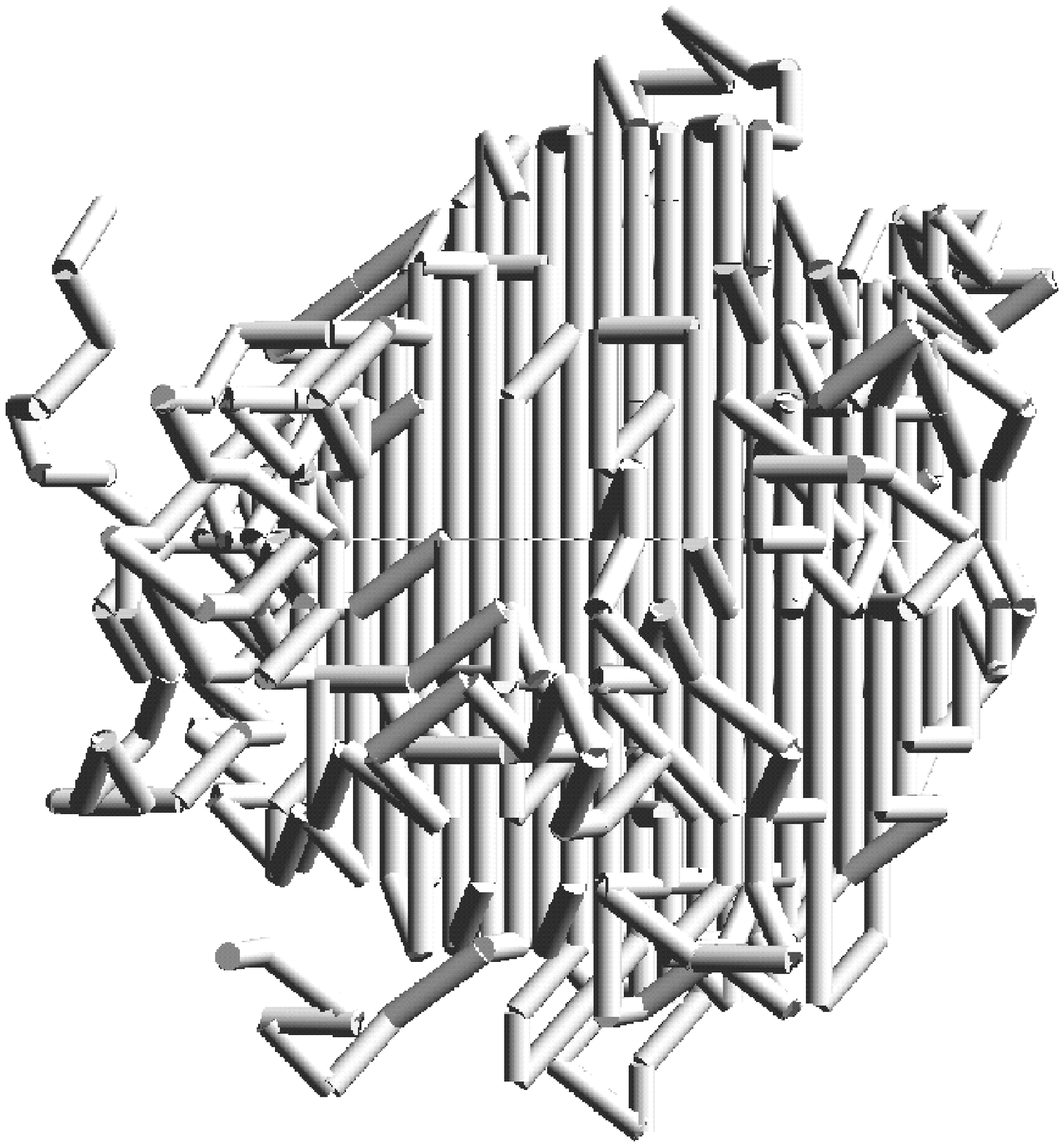,height=6 cm}}
\caption{Radial distributions (3A) of the local crystalline order of the bonds
(see text) for a series of $k_BT/E_p$ values.  The results are averaged over
$5\times10^4$ samples, each with $100$ MC cycles interval. The lines are drawn
as a guide to the eye.  Figure 3B shows a snapshot of single chain at a
temperature $3.086 E_p/k_B$, i.e. just below the irreversible melting
temperature. }
\label{fig3}
\end{figure}

  To follow what is happening, it is useful to trace the structural change on
heating.  To this end, we analyze distribution of local crystalline order as a
function of the distance from the center of mass.  We can quantify the degree of
local crystalline order by counting the number of parallel nearest-neighbor
bonds around each bond (excluding two adjacent bonds along the chain).  In a
dense, ordered state, the number of parallel neighbors is equal to $24$.  In the
coil state, it is more than an order of magnitude lower.  In figure \ref{fig3}A,
we show the local crystalline order as a function of the distance to the center
of mass of the chain.  The figure shows that, as the temperature increases, the
crystalline core is surrounded by a "corona" of bonds that occur in a disordered
environment typical of a polymer coil.  Within the time scale of $10^7$ MC
cycles, the abrupt melting can be observed at about $3.105E_p/k_B$.  Figure
\ref{fig3}B shows a snapshot of the crystallite at a temperature $3.086E_p/k_B$,
i.e. just below the irreversible melting transition.  The snapshot indeed shows
a crystalline core surrounded by a dilute "gas" of coil bonds.  It should be
stressed that these bonds belong to many different loops.  Hence, the surface
disordering is not limited to the chain ends.

  The local crystalline order in the disordered region and its range, increase
with temperature before the point of instability is approached. These results
should be associated to the increase of molten units with temperature, which
enhances the possibility of parallel packing in the disordered region as well as
its outreach.  The uncertainty of the tails in the figure \ref{fig3}A is due to
the scarcity in statistics.
\\  In addition, we found that as long as abrupt melting does not take place,
the number of molten units can be changed reversibly by varying the temperature.
Such behavior is expected when there is a free-energy barrier that separates the
states with partial surface melting from the fully disordered coil state.

\begin{figure}[h]
\centering\epsfig{file=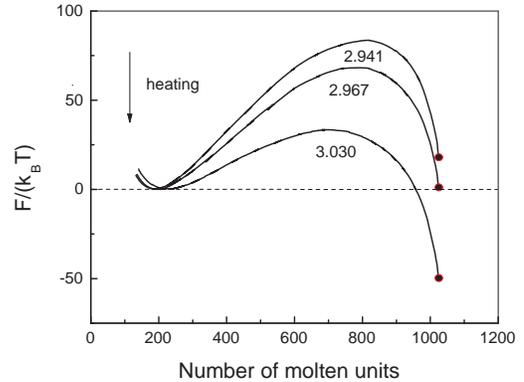,
         height=6 cm}
\caption{Free energy barrier for crystallite melting.  The curves show the
dependence of the free energy on the number of molten polymer units.  The free
energies were calculated using umbrella sampling with $15$ overlapping windows.
Since our estimation is not for absolute free energy, the curves are shifted
longitudinally to meet the first minimum at zero line.
}
\label{fig4}
\end{figure}

   In our simulations, we can directly determine the free-energy barrier for
irreversible melting.  To this end, we compute the probability distribution
$P(n)$ to find $n$ molten units in the system.  The Landau free energy of a
state with this value of $n$ is then given by $F(n)=-k_BTln P(n)$.  To improve
the statistical accuracy of the calculation for those values of $n$ where $P(n)$
is small, we used umbrella sampling. \cite{7} The results of this calculation
are shown in figure \ref{fig4}.  In this figure, we see two minima of the free
energy: one corresponds to the crystallite with a disordered surface, the other
to the disordered coil.  Figure \ref{fig4} allows us to determine the
temperature at which coil and crystalline state are in equilibrium.  At this
temperature of about $2.97 E_p/k_B$, there is a high free-energy barrier (about
$70 k_BT$) separating the crystalline and the molten states.  As the temperature
is increased, the disordered state becomes more stable, but the barrier for
"explosive" melting remains much higher than the thermal energy, up to a reduced
superheating of $4.5 \verb+%+$.  Conversely, as a disordered coil is cooled down
below the coexistence temperature, a large free-energy barrier
will inhibit spontaneous crystallization. In this respect, this
simple homopolymer reproduces some of the aspects of protein
folding.  If we identify the crystalline state with the native
state, and the coil state with the denatured state, then the
present model has a pronounced barrier for both folding and
unfolding. However, we stress that the pathway for melting of
superheated single-chain crystals is very different from the
freezing pathway of supercooled chains.

\textbf{Acknowledgement.} This work was financially supported by DSM Company and
the division of Chemical Science of the Netherlands Organization for Scientific
Research (NWO).  The work of the FOM Institute is part of the research program
of FOM and is made possible by financial support from the NWO.

\end{multicols}


\begin{thebibliography}{30}
\bibitem{1} J. W. M. Frenken and J. F. van der Veen, Phys. Rev. Lett.
\textbf{54},134 (1985); J. G. Dash, Contemp. Phys. \textbf{30}, 89 (1989).
\bibitem{2} L. Grabaek \textit{et al}, Phys. Rev. B \textbf{45}, 2628 (1992); L.
Zhang \textit{et al}, Phys. Rev. Lett. \textbf{85}, 1484 (2000).
\bibitem{3} Z. H. Zhang \textit{et al}, Phys. Rev. B \textbf{57}, 9262 (1998);
X. L. Zeng and H. E.  Elsayed-Ali, \textit{ibid} \textbf{64}, 5410 (2001).
\bibitem{4} A. Keller and S. Z. D. Cheng, Polymer \textbf{39}, 4461 (1998); J.
I. Lauritzen and J. D. Hoffman, J. Res. Nat. Bur. Stds. \textbf{64}, 73 (1960);
D. M. Sadler and G. H. Gilmer, Phys. Rev. Lett. \textbf{56}, 2708 (1986); J. P.
K. Doye and D. Frenkel, J. Chem. Phys. \textbf{109}, 10033 (1998).
\bibitem{5} B. Wunderlich, \textit{Macromolecular Physics vol. 3 crystal
melting} (Academic, New York, 1980).
\bibitem{WM} P.Welch and  M. Muthukumar, Phys. Rev. Lett. 87, 218302(2001).
\bibitem{LM} C. Liu and M. Muthukumar, J. Chem. Phys. 109,
2536(1998)
\bibitem{F} K. Fukui, B.G. Sumpter, M.D. Barnes and D.W. Noid, Comp. Theor.
Pol. SCi.9,245(1999).
\bibitem{TT}A. Trayanov and E. Tosatti Phys. Rev. B 38, 6961-6974
(1988).
\bibitem{6} W. B. Hu, J. Chem. Phys. \textbf{113}, 3901 (2000); \textbf{115},
4395 (2001).
\bibitem{7} G. M. Torrie and J. P. Valleau, Chem. Phys. Lett. \textbf{28}, 578
(1974); S. Auer and D. Frenkel, Nature \textbf{409}, 1020 (2001).


\end{thebibliography}
\end{document}